\documentclass[showpacs,amsmath,amssymb,twocolumn,superscriptaddress,notitlepage,preprintnumbers,pra]{revtex4-1}
\usepackage[dvips]{graphicx}
\usepackage{subfigure}
\usepackage{amsmath,amssymb,amsthm,mathrsfs,amsfonts,dsfont}
\usepackage{dcolumn}
\usepackage{bm}
\usepackage{braket}
\usepackage{physics}
\usepackage{natbib}
\usepackage{mathtools}
\usepackage{diagbox}
\usepackage[utf8]{inputenc}
\usepackage[english]{babel}
\usepackage{scalerel}[2014/03/10]
\usepackage{stackengine}
\usepackage[noend,noline,ruled,linesnumbered]{algorithm2e}
\usepackage[noend]{algpseudocode} 
\usepackage{makecell}
\usepackage{footnotebackref}
\usepackage{lipsum}
\usepackage{hyperref}
\usepackage{multirow}
\usepackage{makecell}
\hypersetup{colorlinks=true, linkcolor=blue, citecolor=blue, urlcolor=black }
\makeatletter

\begin{document}
\title{A Quantum Federated Learning Framework for Classical Clients}
\author{Yanqi Song}
\affiliation{State Key Laboratory of Networking and Switching Technology, Beijing University of Posts and Telecommunications, Beijing, 100876, China
}
\author{Yusen Wu}
\affiliation{
Department of Physics, The University of Western Australia, Perth, WA 6009, Australia
}
\author{Shengyao Wu}
\affiliation{State Key Laboratory of Networking and Switching Technology, Beijing University of Posts and Telecommunications, Beijing, 100876, China
}
\author{Dandan Li}
\affiliation{School of Computer Science (National Pilot Software Engineering School), Beijing University of Posts and Telecommunications, Beijing 100876, China
}
\author{Qiaoyan Wen}
\affiliation{State Key Laboratory of Networking and Switching Technology, Beijing University of Posts and Telecommunications, Beijing, 100876, China
}
\author{Sujuan Qin}
\email{qsujuan@bupt.edu.cn}
\affiliation{State Key Laboratory of Networking and Switching Technology, Beijing University of Posts and Telecommunications, Beijing, 100876, China
}
\author{Fei Gao}
\email{gaof@bupt.edu.cn}
\affiliation{State Key Laboratory of Networking and Switching Technology, Beijing University of Posts and Telecommunications, Beijing, 100876, China
}

\date{\today}

\begin{abstract}
Quantum Federated Learning (QFL) enables collaborative training of a Quantum Machine Learning (QML) model among multiple clients possessing quantum computing capabilities, without the need to share their respective local data. However, the limited availability of quantum computing resources poses a challenge for each client to acquire quantum computing capabilities. This raises a natural question: \emph{Can quantum computing capabilities be deployed on the server instead?} In this paper, we propose a QFL framework specifically designed for classical clients, referred to as CC-QFL, in response to this question. In each iteration, the collaborative training of the QML model is assisted by the shadow tomography technique, eliminating the need for quantum computing capabilities of clients. Specifically, the server constructs a classical representation of the QML model and transmits it to the clients. The clients encode their local data onto observables and use this classical representation to calculate local gradients. These local gradients are then utilized to update the parameters of the QML model. We evaluate the effectiveness of our framework through extensive numerical simulations using handwritten digit images from the MNIST dataset. Our framework provides valuable insights into QFL, particularly in scenarios where quantum computing resources are scarce.
\end{abstract}

\maketitle

\section{Introduction}
Machine learning has made significant progress and brought revolutionary changes across various fields~\cite{simonyan2014very,szegedy2015going,voulodimos2018deep,sutskever2014sequence,silver2016mastering}. However, the reliance of machine learning on sensitive data within the centralized training framework, where data are collected on a single device, poses risks of data leakage. In response, federated learning~\cite{mcmahan2017communication} has emerged as a decentralized training framework that allows multiple devices to collaboratively train a machine learning model without sharing their data. 

In the era of big data, as the scale of data continues to increase, the computational requirements for machine learning are expanding. Simultaneously, theoretical research indicates that quantum computing holds the potential to accelerate the solution of certain problems that pose challenges to classical computers~\cite{harrow2017quantum,shor1999polynomial,grover1997quantum}. Consequently, the field of Quantum Machine Learning (QML)~\cite{biamonte2017quantum,dunjko2018machine,schutt2020machine} has gained widespread attention, with several promising breakthroughs. On one hand, quantum basic linear algebra subroutines, such as Fourier transforms, eigenvector and eigenvalue computations, and linear equation solving, exhibit exponential quantum speedups compared to their well-established classical counterparts~\cite{harrow2009quantum,rebentrost2018quantum,somma2016quantum}. These subroutines bring quantum speedups in a range of machine learning algorithms, including least-squares fitting~\cite{wiebe2012quantum,ref2}, gradient descent~\cite{rebentrost2019quantum,liang2022quantum,ref1,ref3}, principal component analysis~\cite{lloyd2014quantum}, semidefinite programming~\cite{brandao2017quantum}, and support vector machines~\cite{rebentrost2014quantum}. Additionally, another significant development in this field is the Quantum Neural Network (QNN), which is a hybrid quantum-classical machine learning model~\cite{cerezo2021variational,bharti2022noisy}. QNN has demonstrated success in various tasks, including classification~\cite{farhi2018classification,abohashima2020classification,li2022recent,liu2021hybrid,ref4}, regression~\cite{chen2022quantum,kyriienko2021solving,wu2023quantum}, generative learning~\cite{zoufal2019quantum,nakaji2021quantum}, and reinforcement learning~\cite{chen2020variational,lockwood2020reinforcement}.

Quantum Federated Learning (QFL)~\cite{xia2021quantumfed} extends the concept of data privacy preservation into the field of QML. Specifically, the QFL framework consists of a classical server and multiple quantum clients. During each training iteration, the client utilizes its data-encoding quantum states and its QML model to calculate the local gradient. The local gradients of all the clients are then uploaded to the server. The server performs an aggregation process using the received local gradients and shares the global gradient with all clients for updating their model parameters. Since the proposal of QFL, significant research progress has been made in various aspects, including frameworks\cite{chehimi2022quantum,huang2022quantum}, privacy protocols~\cite{xia2021defending,yamany2021oqfl,li2021quantum,zhang2022federated,li2023privacy}, performance optimization techniques~\cite{bhatia2023federated}, and considerations for application-specific scenarios~\cite{zhao2022exact,yun2022slimmable}. However, the scarcity of quantum computing resources poses a challenge for each client to acquire quantum computing capabilities, thereby restricting the applicability of QFL. To overcome this challenge, a solution is to deploy quantum computing capabilities on the server rather than on clients. In this regard, two relevant works were proposed~\cite{sheng2017distributed,li2021quantum}, which enable clients with limited quantum technologies to collaboratively train the QML model on the server. Specifically, Sheng et al.~\cite{sheng2017distributed} proposed a distributed secure quantum machine learning protocol where clients perform single-qubit preparation, operation, and measurement. Additionally, Li et al.~\cite{li2021quantum} proposed a private distributed learning framework based on blind quantum computing~\cite{broadbent2009universal}, where clients prepare single qubits for transmission to the server. Although the preparation of single qubits may not pose significant challenges, ensuring high fidelity in the transmission of qubits through the quantum channel remains a complex issue under current conditions. Thus, the absence of QFL applicable to classical clients emphasizes the necessity for further research in this particular scenario.

In this paper, we propose a QFL framework specifically designed for classical clients, referred to as CC-QFL. In our framework, we encode the local data of each client onto observables instead of quantum states. During the collaborative training of the QML model on the server, we employ the shadow tomography technique~\cite{huang2020predicting,huang2021efficient,nguyen2022optimizing,hadfield2022measurements,wu2022estimating,wu2023overlapped} to remove the necessity of quantum computing capabilities for clients. In detail, the server constructs a classical representation of the QML model, which is then transmitted to the clients. Leveraging this classical representation, clients calculate local gradients based on data-encoding observables, and these local gradients are subsequently utilized to update the parameters of the QML model. We conduct extensive numerical simulations of our framework using handwritten digit images from the MNIST dataset. The excellent numerical performance validates the feasibility and effectiveness of CC-QFL. Our framework contributes to the advancement of QFL, particularly in scenarios where quantum computing resources are limited.

\section{QFL Framework}\label{sec_qfl}
In this section, we present the conventional framework of QFL. To provide a clear description, we begin with a detailed introduction to the QNN model which represents a hybrid quantum-classical machine learning model.

\begin{figure*}[htpb]
\centering
\includegraphics[width=0.99\textwidth]{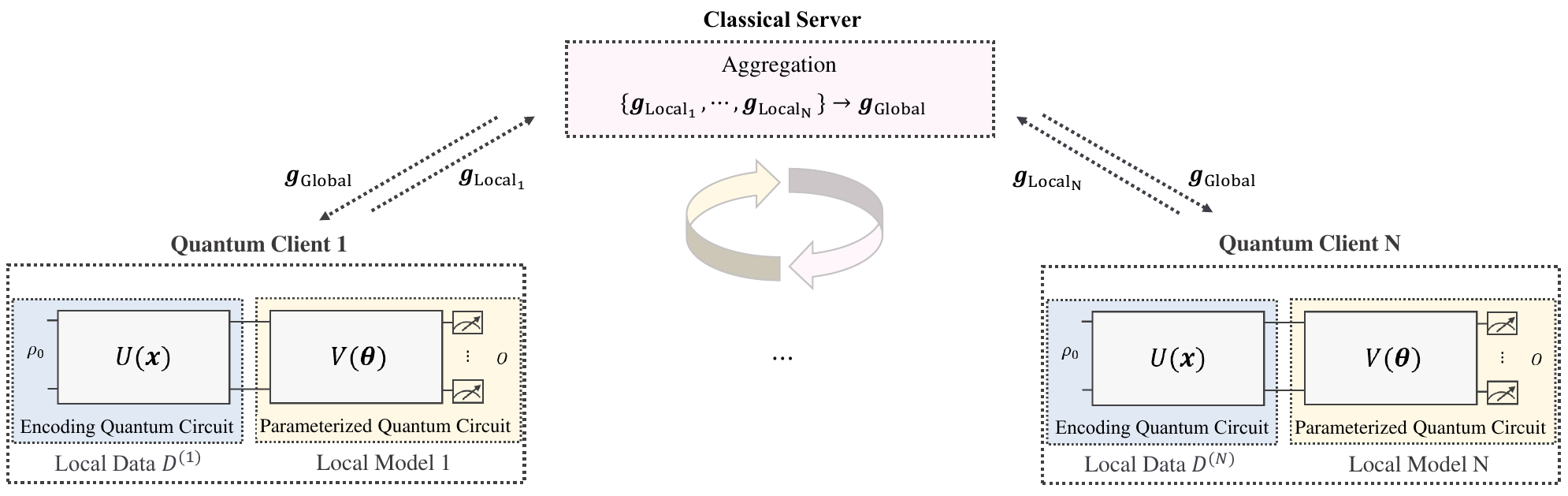} 
\caption{\textbf{Illustration of QFL framework}. 
In each training iteration, each quantum client $i$ encodes its local data $D^{(i)}$ onto quantum states and utilizes its parameterized quantum circuit $V(\bm \theta)$ to calculate the local gradient $\bm{g}_{\rm{Local_i}}$. These local gradients $\{\bm{g}_{\rm{Local_i}}\}_{i=1}^N$ are then uploaded to the classical server. The server aggregates $\{\bm{g}_{\rm{Local_i}}\}_{i=1}^N$ to obtain the global gradient $\bm{g}_{\rm{Global}}$ and subsequently shares $\bm{g}_{\rm{Global}}$ with all clients. Each client then updates its model parameters $\bm\theta$.}
\label{fig_1}
\end{figure*}

\subsection{Structure of QNN Model}\label{sec_2_1}
The QNN model begins by encoding the classical data $\bm x \in \mathbb{R}^d$ onto a quantum state $\rho(\bm x)$. This data encoding process involves applying an encoding quantum circuit $U(\bm x)$ to an initial state $\rho_0$. Following the data encoding process, a parameterized quantum circuit $V(\bm\theta)$ is applied to $\rho(\bm x)$, where the model parameters $\bm\theta \in \mathbb{R}^p$ can be updated during the training process. Afterward, the resulting state is measured with respect to an observable $O$, yielding the following parameterized expectation value
\begin{equation}\label{eq_1}
    E(\bm \theta)={\rm Tr}[V(\bm \theta)\rho(\bm x)V^{\dagger}(\bm \theta)O].
\end{equation}
Here, $\rho(\bm x)=U(\bm x)\rho_0U^{\dagger}(\bm x)$. Now, $E(\bm{\theta})$ can be used to calculate a loss function $\mathcal{L}(\bm{\theta})$ that quantifies the difference between the predicted output of the QNN model and the expected output. The selection of $\mathcal{L}(\bm{\theta})$ depends on the task type. For regression tasks, mean squared error is commonly used, while for classification tasks, cross-entropy loss is often employed. In training the QNN model, classical optimization methods such as Stochastic Gradient Descent (SGD)~\cite{kingma2014adam} are employed to iteratively update $\bm{\theta}$ to optimize $\mathcal{L}(\bm{\theta})$.

\subsection{Framework of QFL}
Now, we provide a detailed introduction to QFL which comprises a classical server and $N$ quantum clients. Given that the $i$-th client holds the local data
\begin{equation}\label{eq_2}
     D^{(i)}=\{\bm x^{(i)}_j\}_{j=1}^{m_i},
\end{equation}
where $\bm x^{(i)}_j\in \mathbb{R}^d$, and $m_i$ represents the number of data held by the $i$-th client. In each training iteration, each client $i$ encodes its local data $D^{(i)}$ onto quantum states 
\begin{equation}
    \rho^{(i)}=\{\rho(\bm x_{j}^{(i)})\}_{j=1}^{m_i},
\end{equation}
where $\rho(\bm x_{j}^{(i)})=U(\bm x_{j}^{(i)})\rho_0U^{\dagger}(\bm x_{j}^{(i)})$. Next, each client $i$ utilizes its data-encoding quantum states $\rho^{(i)}$ and its parameterized quantum circuit $V(\bm\theta)$ to calculate the local gradient $\bm{g}_{\rm{Local_i}}$ based on the specific form of the loss function $\mathcal{L}(\bm\theta)$. The local gradients of all the clients $\{\bm{g}_{\rm{Local_i}}\}_{i=1}^N$ are then uploaded to the server. The server performs an aggregation process using $\{\bm{g}_{\rm{Local_i}}\}_{i=1}^N$ and obtains the global gradient
\begin{equation}\label{eq_3}
\bm{g}_{\rm{Global}}=\sum_{i=1}^{N}w_i \bm{g}_{\rm{Local_i}}.
\end{equation}
In the natural setting, the weight $w_i=m_i/m$, where $m$ represents the total number of data from all clients. The server shares $\bm{g}_{\rm{Global}}$ with all clients, allowing each client to update its corresponding model parameters as follows
\begin{equation}\label{eq_4}
    \bm{\theta} \leftarrow\bm{\theta}-\eta\bm{g}_{\rm{Global}}.
\end{equation}
Here, $\eta$ represents the learning rate. The process of calculating local gradients, aggregating these gradients, and updating the model parameters on each client is repeated iteratively until a predefined number of iterations is reached. The complete framework of QFL is illustrated in Fig.~\ref{fig_1}.

However, the limited availability of quantum computing resources presents a challenge as clients may struggle to acquire their quantum computing capabilities, consequently restricting the practicality of QFL. To address this challenge, it is worth considering a paradigm shift by placing quantum computing capabilities on the server instead of the clients. In the following section, we will provide a comprehensive description of our QFL framework specifically designed for classical clients, referred to as CC-QFL. 

\section{Main Results: CC-QFL Framework}
As a fundamental component of the QFL framework, we first introduce our QNN model.
\subsection{Structure of Our QNN Model}
In our QNN model, we directly apply a parameterized quantum circuit $V(\bm\theta)$ to an initial state $\rho_0$ to obtain a parameterized quantum state $\rho(\bm \theta)$, where the model parameters $\bm\theta \in \mathbb{R}^p$ can be updated during the training process. Subsequently, we measure $\rho(\bm \theta)$ with respect to a data-encoding observable $O(\bm x)$, where the classical data $\bm x \in \mathbb{R}^d$, resulting in the following parameterized expectation value
\begin{equation}\label{eq_5}
    \Tilde{E}(\bm {\theta})={\rm Tr}[\rho(\bm \theta)O(\bm x)].
\end{equation}
Here, $\rho(\bm \theta)=V(\bm \theta)\rho_0V^{\dagger}(\bm \theta)$, and  $O(\bm x)=\sum_{j=1}^{l}c_j(\bm x)O_j$. The coefficients $\{c_j(\bm x)\}_{j=1}^l$ associated with the Hermitian matrices $\{O_j\}_{j=1}^l$ are obtained from the data-dependent function $c(\bm x):\mathbb{R}^d \rightarrow \mathbb{R}^l$. Now, $\Tilde{E}(\bm{\theta})$ can be used to calculate a loss function $\Tilde{\mathcal{L}}(\bm{\theta})$, and classical optimization methods such as SGD can be employed to iteratively update $\bm\theta$ during the training process to optimize $\Tilde{\mathcal{L}}(\bm{\theta})$. In detail, according to the chain rule, the derivatives $\{\partial \Tilde{\mathcal{L}}(\bm{\theta})/\partial \theta_k\}_{k=1}^p$ can be calculated using $\Tilde{E}(\bm {\theta})$ and the derivatives $\{\partial \Tilde{E}(\bm {\theta})/\partial \theta_k\}_{k=1}^p$. Furthermore, due to the ``parameter shift rule"~\cite{mitarai2018quantum,schuld2019evaluating}, $\partial \Tilde{E}(\bm {\theta})/\partial \theta_k$ can be expressed as
\begin{equation}\label{eq_6}
\partial \Tilde{E}(\bm {\theta})/\partial \theta_k = [ \Tilde{E}_{\theta_k+}(\bm {\theta})- \Tilde{E}_{\theta_k-}(\bm {\theta})]/2,
\end{equation}
where $\Tilde{E}_{\theta_k\pm}(\bm {\theta})$ denotes $\Tilde{E}(\bm {\theta})$ with $\theta_k$ being $\theta_k \pm \pi/2$. In summary, calculating $\{\partial \Tilde{\mathcal{L}}(\bm{\theta})/\partial \theta_k\}_{k=1}^p$ can be transformed into calculating the following expectation values
\begin{equation}\label{eq_7}
    \{\Tilde{E}(\bm {\theta}),\Tilde{E}_{\theta_k\pm}(\bm {\theta})\}_{k=1}^p.
\end{equation}

A notable characteristic of our QNN model is the encoding of classical data $\bm x$ onto the observable $O(\bm x)$ instead of the quantum state $\rho(\bm x)$. This characteristic distinguishes our model from conventional QNN models as described in Sec.~\ref{sec_2_1}. 

\begin{algorithm*}[t]
\caption{\textbf{Training procedure of CC-QFL}}\label{alg_cc-QFL}
\DontPrintSemicolon
\SetKwInOut{Input}{Input}
\SetKwInOut{Output}{Output}
\SetKwInOut{Initialize}{Initialize}
\SetKwFor{While}{while}{do}{}
\SetKwIF{If}{ElseIf}{Else}{if}{then}{else if}{else}{end if}
\Input{the initial state $\rho_0$, the parameterized quantum circuit $V(\bm\theta)$ with $\bm\theta \in \mathbb{R}^p$, the number of measurements $M$, the learning rate $\eta$, the number of clients $N$, the number of data held by the $i$-th client $m_i$, the loss function $\Tilde{\mathcal{L}}(\bm \theta)$, and the number of iterations $T$}
\Output {the trained model parameters $\bm \theta^{(T)}$}
\Initialize {the random model parameters $\bm\theta^{(1)}$}
\While{$1 \leq t\leq T$}{
\qquad
Construct the classical shadows $\mathcal{S}_{\rm Global}^{(t)}$~(Eq.~\ref{eq_11}) of the parameterized quantum states $\{(\rho(\bm \theta^{(t)}),\rho_{\theta_k\pm}(\bm \theta^{(t)})\}_{k=1}^p$ by \;
\qquad the server, where $\rho(\bm \theta^{(t)})=V(\bm \theta^{(t)})\rho_0V^{\dagger}(\bm \theta^{(t)})$, and $\rho_{\theta_k\pm}(\bm \theta^{(t)})$ denotes $\rho(\bm \theta^{(t)})$ with $\theta_k$ being $\theta_k \pm \pi/2$ \;
\qquad Transmit $\mathcal{S}_{\rm Global}^{(t)}$ to clients \;
\qquad \While{$1 \leq i\leq N$}{ 
\qquad\qquad
Calculate local gradient $\bm{g}_{\rm{Local_i}}^{(t)}$ by the $i$-th client, leveraging $\mathcal{S}_{\rm Global}^{(t)}$, data-encoding observables $O^{(i)}$~(Eq.~\ref{eq_8}),\;
\qquad \qquad and the specific form of the loss function $\Tilde{\mathcal{L}}(\bm \theta^{(t)})$\;
\qquad\qquad Set $i\gets i+1$}
 \qquad Upload local gradients of all clients $\{\bm{g}_{\rm{Local_i}}^{(t)}\}_{i=1}^N$ to the server\; 
 \qquad Aggregate $\{\bm{g}_{\rm{Local_i}}^{(t)}\}_{i=1}^N$ as described in Eq.~\ref{eq_3} to obtain the global gradient $\bm{g}_{\rm{Global}}^{(t)}$ by the server\;
 \qquad Update model parameters $\bm \theta^{(t)}$ by the server according to Eq.~\ref{eq_4}\;
\qquad Set $t\gets t+1$}
\Return the trained model parameters $\bm\theta^{(T)}$
\end{algorithm*}

\subsection{Framework of CC-QFL}
The significance of the aforementioned characteristic becomes particularly evident in scenarios where $N$ classical clients collaboratively train a QNN model, leveraging the quantum computing capabilities of the server, without the need to share their respective local data. In this context, each client $i$ encodes its local data $D^{(i)}$~(Eq.~\ref{eq_2}) onto observables 
\begin{equation}\label{eq_8}
    O^{(i)}=\{O(\bm x_j^{(i)})\}_{j=1}^{m_i},
\end{equation}
where $O(\bm x_j^{(i)})=\sum_{h=1}^{l}c_h(\bm x_j^{(i)})O_h$. The initial state $\rho_0$ and the parameterized quantum circuit $V(\bm \theta)$ are deployed on the server. Furthermore, in order to remove the necessity of quantum computing capabilities for clients, the collaborative training of the QNN model is assisted by the shadow tomography technique~\cite{huang2020predicting}. This technique constructs the classical representation of the given state to evaluate the expectation values of certain observables classically. A more detailed description of the shadow tomography technique is provided in Appendix~\ref{appendix_1}.

The collaborative training of the QNN model consists of two important phases: (i)~an acquisition phase where the server constructs a classical representation of the parameterized quantum states $\{(\rho(\bm \theta),\rho_{\theta_k\pm}(\bm \theta)\}_{k=1}^p$, where $\rho(\bm \theta)=V(\bm \theta)\rho_0V^{\dagger}(\bm \theta)$, and $\rho_{\theta_k\pm}(\bm \theta)$ denotes $\rho(\bm \theta)$ with $\theta_k$ being $\theta_k \pm \pi/2$. (ii)~an evaluation phase where clients leverage this classical representation and data-encoding observables of all clients $\{O^{(i)}\}_{i=1}^N$ to calculate local gradients $\{\bm{g}_{\rm{Local_i}}^{(t)}\}_{i=1}^N$ for updating the model parameters $\bm \theta$. The complete training procedure of CC-QFL is described in Alg.~\ref{alg_cc-QFL}.

During the acquisition phase, the server repeatedly executes a simple measurement procedure. This procedure involves randomly selecting a unitary operator $W$ from a fixed ensemble $\mathcal{W}$ to rotate the $n$-qubit parameterized quantum state $\rho(\bm{\theta})$, followed by a computational-basis measurement. Upon receiving the $n$-bit measurement outcome $|b(\bm \theta)\rangle \in \{0,1\}^n$, a completely classical post-processing step applies the inverted channel $\mathcal{M}^{-1}$ to $W^\dagger |b(\bm\theta)\rangle \langle b(\bm\theta)| W$, where $\mathcal{M}$ depends on the ensemble $\mathcal{W}$. Consequently, a classical snapshot $\hat{\rho}(\bm \theta)$ of $\rho(\bm \theta)$ is obtained from a single measurement, given by
\begin{equation}\label{eq_9}
    \hat{\rho}(\bm \theta)=\mathcal{M}^{-1}(W^\dagger |b(\bm\theta)\rangle \langle b(\bm\theta)| W).
\end{equation}
By repeating the aforementioned procedure $M$ times, the classical shadow~(classical representation) $S(\rho(\bm \theta); M)$ of $\rho(\bm{\theta})$ is obtained and defined as
\begin{equation}\label{eq_10}
S(\rho(\bm \theta); M)=\{\hat{\rho}_{j}(\bm \theta)\}_{j=1}^{M},
\end{equation}
where $\hat{\rho}_{j}(\bm \theta)=\mathcal{M}^{-1}(W_j^\dagger |b_j(\bm \theta)\rangle \langle b_j(\bm \theta)| W_j)$. The server ultimately constructs the classical shadows of parameterized quantum states $\{(\rho(\bm \theta),\rho_{\theta_k\pm}(\bm \theta))\}_{k=1}^p$. These classical shadows are expressed as
\begin{equation}\label{eq_11}
   \mathcal{S}_{\rm Global}=
   \{(S(\rho(\bm \theta); M),S(\rho_{\theta_k\pm}(\bm \theta); M)\}_{k=1}^p.
\end{equation}
Then, $\mathcal{S}_{\rm Global}$ are transmitted to clients.

\begin{figure*}[ht]
\centering
\includegraphics[width=0.99\textwidth]{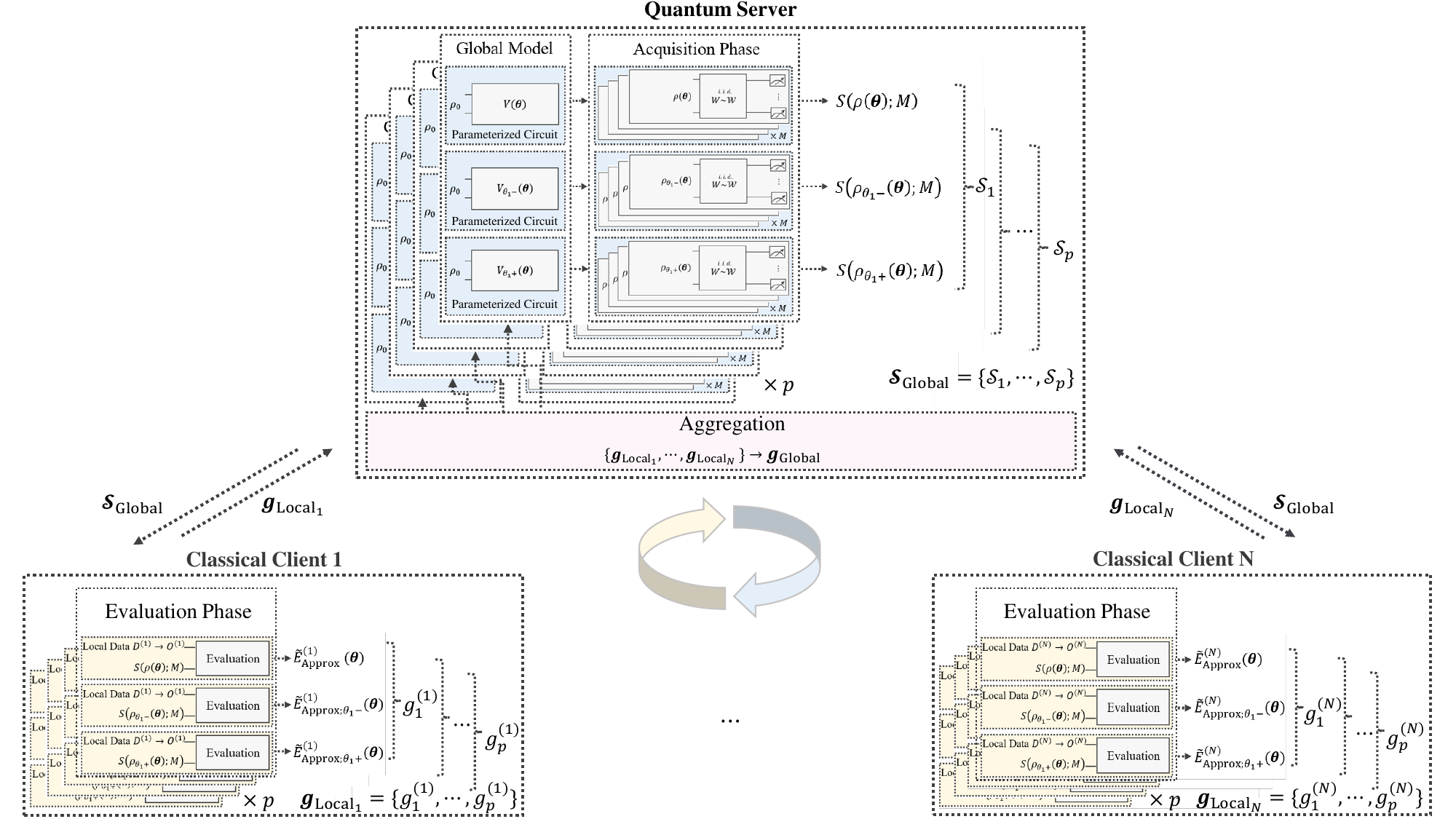} 
\caption{\textbf{Illustration of CC-QFL framework}. Each classical client $i$ encodes its local data $D^{(i)}$ onto observables $O^{(i)}$, while the initial state $\rho_0$ and the parameterized quantum circuit $V(\bm \theta)$ are deployed on the quantum server. In each training iteration, the server constructs the classical shadows $\mathcal{S}_{\rm Global}$ of the parameterized quantum states $\{(\rho(\bm \theta),\rho_{\theta_k\pm}(\bm \theta))\}_{k=1}^p$. $\mathcal{S}_{\rm Global}$ are then transmitted to the clients. Each client $i$ leverages $\mathcal{S}_{\rm Global}$ and its data-encoding observables $O^{(i)}$ to calculate the corresponding local gradient $\bm{g}_{\rm{Local_i}}$. The local gradients of all clients $\{\bm{g}_{\rm{Local_i}}\}_{i=1}^N$ are uploaded to the server. The server performs an aggregation process using $\{\bm{g}_{\rm{Local_i}}\}_{i=1}^N$ to obtain the aggregated global gradient $\bm{g}_{\rm{Global}}$. Finally, the server updates its model parameters $\bm \theta$.}
\label{fig_2}
\end{figure*}

During the evaluation phase, each client $i$ leverages $\mathcal{S}_{\rm Global}$ and its data-encoding observables $O^{(i)}$ to evaluate the expectation values $\{\Tilde{E}^{(i)}(\bm {\theta}),\Tilde{E}^{(i)}_{\theta_k\pm}(\bm {\theta})\}_{k=1}^p$~(Eq.~\ref{eq_7}) using median of means estimation as described in Alg.~\ref{algorithm B1}. This estimation method provides the approximations $\{\Tilde{E}_{\rm Approx}^{(i)}(\bm {\theta}),\Tilde{E}^{(i)}_{\rm Approx;\theta_k\pm}(\bm {\theta})\}_{k=1}^p$ of $\{\Tilde{E}^{(i)}(\bm {\theta}),\Tilde{E}^{(i)}_{\theta_k\pm}(\bm {\theta})\}_{k=1}^p$, which are used to calculate the corresponding local gradient $\bm{g}_{\rm{Local_i}}$ based on the specific form of the loss function $\Tilde{\mathcal{L}}(\bm \theta)$. 

Subsequently, the local gradients of all clients $\{\bm{g}_{\rm{Local_i}}\}_{i=1}^N$ are uploaded to the server. The server performs an aggregation process using $\{\bm{g}_{\rm{Local_i}}\}_{i=1}^N$ to obtain the aggregated global gradient $\bm{g}_{\rm{Global}}$ as described in Eq.~\ref{eq_3}. Finally, the server updates its model parameters $\bm \theta$ according to Eq.~\ref{eq_4}. The process of constructing classical shadows, calculating local gradients, aggregating these gradients, and updating the model parameters on the server is repeated iteratively until a predefined number of iterations is reached.

In summary, our framework allows multiple classical clients to collaborate in training a QNN model.
By encoding the local data of each client onto observables and employing the shadow tomography technique, the clients can actively participate in the training process without the need to share their local data while leveraging the quantum computing capabilities of the server. The complete framework of CC-QFL is depicted in Fig.~\ref{fig_2}.

\section{Experiment Results}\label{sec:Theoretical Analysis 1} 
We perform numerical simulations of our CC-QFL framework using the TensorFlow Quantum~\cite{broughton2020tensorflow} simulation platform. These simulations employ handwritten digit images from the MNIST dataset, which comprises 70,000 images. The experimental settings are summarized as follows.

\begin{figure*}[htpb]
    \centering
    \subfigure[]{
		\begin{minipage}[b]{0.45\textwidth}
			\includegraphics[width=1\textwidth]{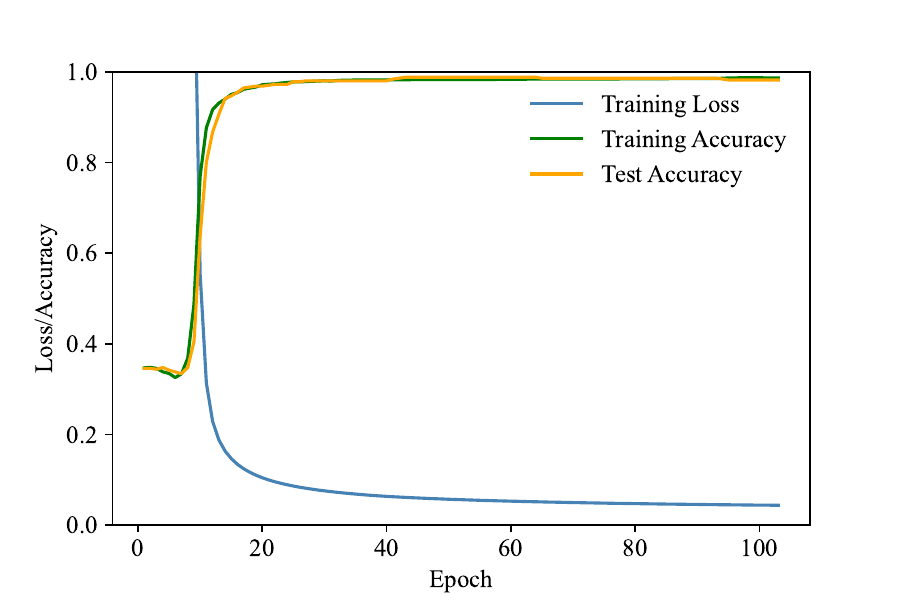}
		\end{minipage}
		\label{fig_3:subfig1}
    }
    \subfigure[]{
		\begin{minipage}[b]{0.45\textwidth}
			\includegraphics[width=1\textwidth]{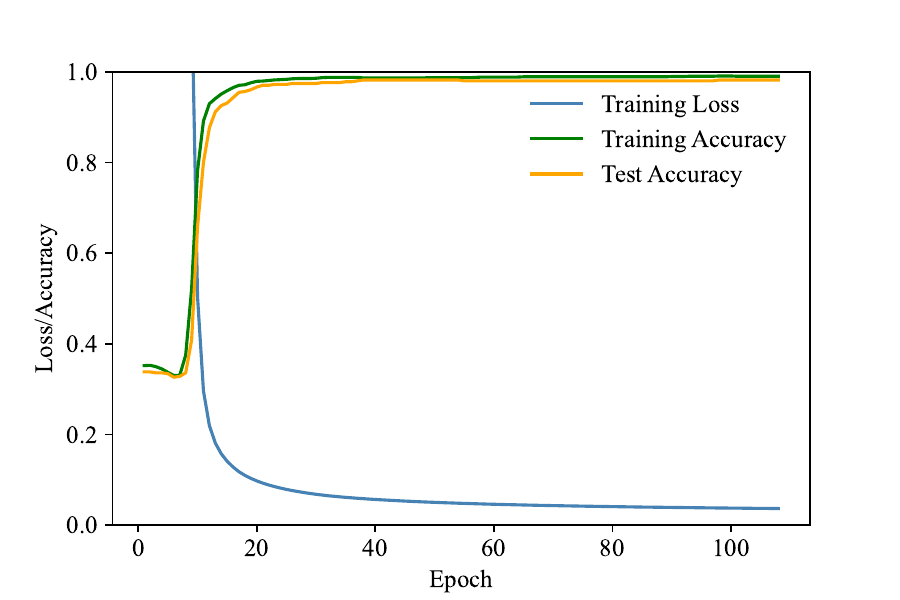}
		\end{minipage}
		\label{fig_3:subfig2}
    }
    \caption{\textbf{Performance of CC-QFL in the binary classification task on the MNIST dataset.} (a) the numerical results of the single-client scenario. (b) the numerical results of the multi-client scenario ($N=3$). Here, the training loss, training accuracy, and test accuracy are represented by solid lines in blue, green, and orange, respectively.}
    \label{fig_3}
\end{figure*}

First, we preprocess the handwritten digit images. Each image is flattened into a one-dimensional vector of length 784, representing its 28x28 pixels. Due to the computational limitations of the simulation, we normalize the image and perform dimensionality reduction using the principal component analysis algorithm~\cite{ma1993principal}. This reduces the vector length to 8. The resulting image is denoted as $\bm x=[x_1,\cdots,x_8]^{T}$. The classical clients encode $\bm x$ onto the observable $O(\bm x)=\sum_{j=1}^8x_{j}\sigma ^z_j$, where $\sigma ^z_j$ is the Pauli $Z$ operator acting on the $j$-th qubit. The parameterized quantum circuit is a 5-layer hardware efficient ansatz~\cite{kandala2017hardware}, which is deployed on the quantum server. To construct the corresponding classical shadows, the server measures each qubit independently in a random Pauli basis. 

In the subsequent sections, we will present the performance of single-client and multi-client CC-QFL frameworks in a binary classification task. To achieve this, we employ the cross-entropy loss function.

\subsection{Performance of Single-client CC-QFL}
In the single-client CC-QFL framework, a classical client utilizes the quantum computing capabilities of the server to train its classifier model without the need to upload its local data to the server. The classification task involves accurately classifying handwritten digit images of the numbers ``3" and ``6" from the MNIST dataset. The client has a total of 2640 images, with 2000 images allocated for the training set and the remaining 640 images reserved as the test set. The specific parameter settings of the experiment can be found in Tab.~\ref{hyperp}.

The training and testing results of our numerical simulations are plotted in Fig.~\ref{fig_3:subfig1}. During the first 20 epochs, it is observed that the training loss exhibits a rapid decrease, while both the training accuracy and test accuracy show a rapid increase following a slight decline. Upon reaching convergence, the training accuracy and test accuracy reached $98.69\%$ and $98.24\%$ respectively. These results demonstrate that the single-client CC-QFL achieves excellent numerical performance in the binary classification task on the MNIST dataset.

\begin{table}[h]
\caption{\textbf{Parameter settings for single-client and multi-client CC-QFL.}}
\begin{tabular}{lcc}
  \hline
  Parameter & Single-client  & Multi-client   \\
  \hline
  num of clients    & 1  & 3 \\
  num of classes   & 2 (``3" or ``6") & 2 (``3" or ``6")   \\
  num of qubits    & 8 & 8  \\
  circuit depth   & 5 & 5 \\
  data distribution  & /   & non-i.i.d  \\
  training set (per client)   & 2000 & 700  \\
  test set (per client)  & 640 & 220  \\
  optimizer   & Adam & Adam  \\
  learning rate   & 0.003 & 0.003  \\
  loss function   & cross-entropy & cross-entropy  \\
  \hline
\end{tabular}
\label{hyperp}
\end{table}

\subsection{Performance of Multi-client CC-QFL}
\begin{figure}[h]
\centering
\includegraphics[width=0.48\textwidth]{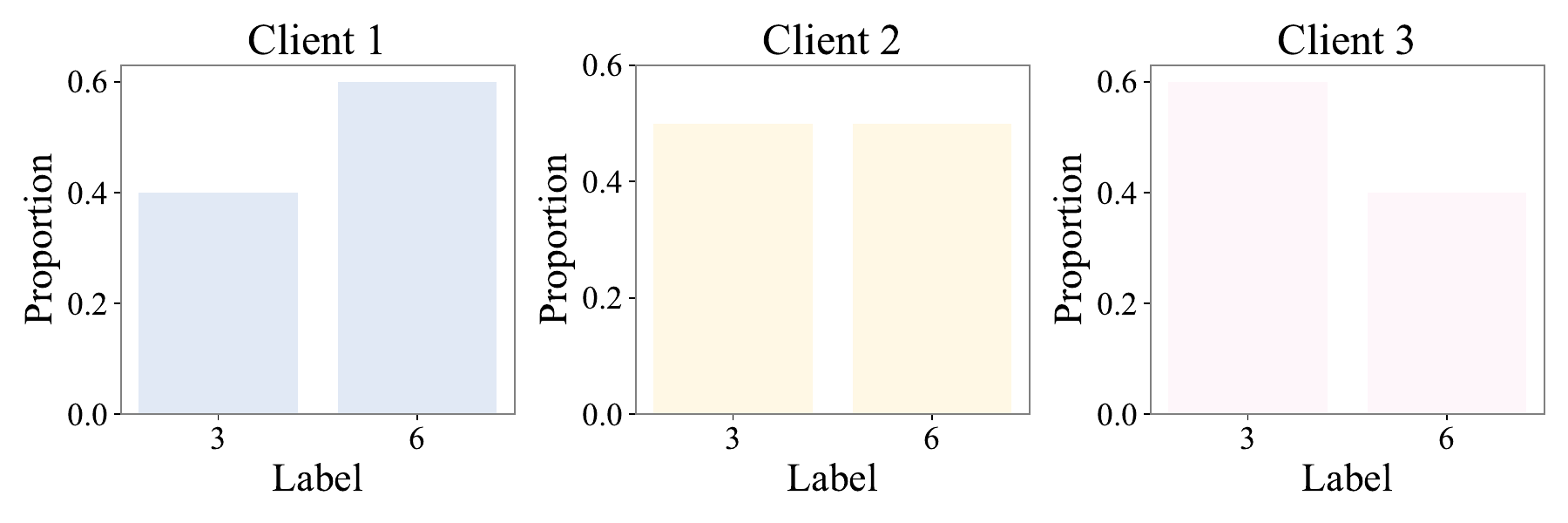} 
\caption{\textbf{Heterogeneity of the training data.} We illustrate the proportion of handwritten digit labels (``3" and ``6") on each client, where each client contains a total of 700 images.}
\label{fig_4}
\end{figure}

Compared to the single-client scenario, our primary focus lies in the performance of the multi-client CC-QFL, where $N$ classical clients leverage the quantum computing capabilities of the server to collaboratively train the classifier model, without the need to upload their respective local data to the server. In this scenario, with N=3, the classification task focuses on accurately classifying handwritten digit images of the numbers ``3" and ``6" from the MNIST dataset. Each client is assigned a total of 920 images, with 700 images allocated for the training set and the remaining 220 images reserved for the test set. Considering the uniqueness of each client in real-world scenarios, the training data should satisfy non-i.i.d. To address this, we utilize Leaf~\cite{caldas2018leaf} to process the MNIST dataset, ensuring that each client has a distinct proportion of handwritten digit labels. To visualize the non-i.i.d nature of the training data, the corresponding heterogeneity is depicted in Fig.~\ref{fig_4}. The specific parameter settings of the experiment can be found in Tab.~\ref{hyperp}.

The training and testing results obtained from our numerical simulations are illustrated in Fig.~\ref{fig_3:subfig2}. The numerical results reveal a significant decrease in training loss during the first 20 epochs. Furthermore, both the training accuracy and test accuracy demonstrate a notable increase around the 10th epoch. At convergence, the training accuracy achieves $98.69\%$, and the test accuracy reaches $98.05\%$, which is comparable to the numerical results achieved in the single-client scenario. The excellent numerical performance validates the feasibility and effectiveness of multi-client CC-QFL.

\section{Conclusion}
In this paper, we present a novel QFL framework designed to address scenarios involving limited quantum computing resources. 
This framework allows classical clients to collaboratively train a QML model while ensuring the privacy of their data. In contrast to the conventional QFL framework, our framework involves deploying the QML model on the server instead of the clients and encoding the classical data onto observables rather than quantum states. Furthermore, the shadow tomography technique assists in the training of the QML model and eliminates the need for clients to possess quantum computing capabilities in the conventional QFL framework. Our framework extends the potential applications of QFL, particularly in situations where quantum computing resources are scarce.

Furthermore, there remain several intriguing aspects that require further investigation. One aspect of the investigation involves exploring different approaches for encoding classical data onto observables and analyzing their impact on the effectiveness of the QML model training to derive more appropriate data encoding approaches. An additional investigation aspect involves mitigating the potential leakage of client data to the server through gradient inversion attacks~\cite{zhu2019deep,geiping2020inverting}. In this regard, the adoption of secure gradient aggregation strategies is crucial. Our framework can naturally integrate with existing techniques such as homomorphic encryption~\cite{gentry2009fully} or differential privacy~\cite{dwork2008differential,dwork2006differential} to enhance client data security. Moreover, the exploration of novel strategies holds significant value.

\section*{Acknowledgments}
This work is supported by National Natural Science Foundation of China (Grant Nos.~62272056, 62372048, 62371069), Beijing Natural Science Foundation (Grant No.~4222031), and China Scholarship Council (Grant No.~202006470011).

\bibliography{Quantum_Federated_Learning}

\clearpage
\widetext
\widetext
\appendix

\section{Shadow Tomography Technique}\label{appendix_1}
The shadow tomography technique, proposed by Huang et al.~\cite{huang2020predicting}, is a method used to extract meaningful information from an $n$-qubit unknown quantum state $\rho$. One important application of this technique is to construct the classical shadow of $\rho$ in order to evaluate the expectation values $\{o_i\}_{i=1}^H$ of certain observables $\{O_{i}\}_{i=1}^{H}$ classically. Here, $o_i=\mathrm{tr}(O_i\rho)$.

\begin{algorithm*}[t]
\caption{\textbf{Shadow tomography technique for evaluating expectation values}}\label{algorithm B1}
\DontPrintSemicolon
\SetKwInOut{Input}{Input}
\SetKwInOut{Output}{Output}
\SetKwInOut{Initialize}{Initialize}
\SetKwFor{While}{while}{do}{}
\SetKwIF{If}{ElseIf}{Else}{if}{then}{else if}{else}{end if}
\textbf{Stage 1: Construct classical shadow}\;
\Input{the $n$-qubits unknown state $\rho$, the set of unitary operators $\mathcal{W}$, and the number of measurements $M$}
\Output {the classical shadow $S\left(\rho;M\right)=\left\{ \hat{\rho}_1, \cdots,  \hat{\rho}_M \right\}$}
\While{$1 \leq j \leq M$}{
\qquad Randomly select a unitary operator $W_j$ from $\mathcal{W}$ and apply $W_j$ to $\rho$\;
\qquad Measure in the computational basis and obtain $b_j\in \{ 0,1\}^n$\;
\qquad Apply the inverted quantum channel $\mathcal{M}^{-1}$ to $W_j^\dagger |b_j\rangle \langle b_j| W_j$ and obtain $\hat{\rho}_j= \mathcal{M}^{-1} (W_j^\dagger  | b_j \rangle \langle b_j  | W_j )$, where $\mathcal{M}$ \;
\qquad depends on the set $\mathcal{W}$}
\Return the classical shadow $S\left(\rho;M\right)=\left\{ \hat{\rho}_1, \cdots,  \hat{\rho}_M \right\}$\;
\textbf{Stage 2: Evaluate expectation values} \; 
\Input{the set of observables $\{ O_i \}_{i=1}^H$, and the classical shadow $S\left(\rho;M\right)=\left\{ \hat{\rho}_1, \cdots,  \hat{\rho}_M \right\}$}
\Output {the approximations $\{\Tilde{o}_{i}\}_{i=1}^H$ of the expectation values $\{o_i\}_{i=1}^H$, where $o_i=\mathrm{tr}(O_i\rho)$}
Divide $S\left(\rho;M\right)$ into $M_2$ equally sized chunks $\{ \hat{\rho}_1^{(1)}, \cdots,  \hat{\rho}_{M_1}^{(1)} \},\cdots, \{ \hat{\rho}_1^{(M_2)}, \cdots,  \hat{\rho}_{M_1}^{(M_2)} \}$, where $M=M_1 \times M_2$\;
\While{$1\leq i \leq H$}{
\qquad \While{$1 \leq k \leq M_2$}{
\qquad \qquad Evaluate $o_i$ using the $k$-th chunk and obtain the $k$-th approximation $\tilde{o}_i^{(k)}=\frac{1}{M_1}\sum_{l=1}^{M_1}\mathrm{tr}(O_i \hat{\rho}_{l}^{(k)} )$
}
\qquad Calculate the median of the $M_2$ approximations and obtain the final approximation $\tilde{o}_i=\mathrm{median}\left\{ \tilde{o}_i^{(1)}, \cdots, \tilde{o}_i^{(M_2)}\right \}$
}
\Return the approximations $\{\Tilde{o}_{i}\}_{i=1}^H$
\end{algorithm*}

Specifically, a simple measurement procedure is repeatedly executed to construct the classical shadow of $\rho$. This procedure involves randomly selecting a unitary operator $W$ from a fixed ensemble $\mathcal{W}$ to rotate $\rho$, followed by a computational-basis measurement. Upon receiving the $n$-bit measurement outcome $|b\rangle \in \{0,1\}^n$, the classical memory can efficiently store a classical description of $W^\dagger |b\rangle \langle b| W$, according to the Gottesman-Knill theorem~\cite{gottesman1997stabilizer}. It is instructive to consider the average mapping from $\rho$ to $W^\dagger |b\rangle \langle b| W$ as a quantum channel given by
\begin{equation}
\mathcal{M}(\rho)=\mathbb{E}[W^\dagger |b\rangle \langle b| W],
\end{equation}
where $\mathcal{M}$ depends on the ensemble $\mathcal{W}$. Consequently, a completely classical post-processing step applies the inverted quantum channel $\mathcal{M}^{-1}$ to the measurement outcome $W^\dagger |b\rangle \langle b| W$. Subsequently, a classical snapshot $\hat{\rho}$ of $\rho$ is obtained from a single measurement, given by
\begin{equation}
  \hat{\rho}=\mathcal{M}^{-1}(W^\dagger |b\rangle \langle b| W). 
\end{equation}
By repeating the aforementioned procedure $M$ times, the classical shadow of $\rho$ is obtained and defined as
\begin{equation}
S(\rho; M)=\{\hat{\rho}_{j}\}_{j=1}^{M},
\end{equation}
where $\hat{\rho}_{j}=\mathcal{M}^{-1}(W_{j}^\dagger |b_j\rangle \langle b_j| W_j)$. $S(\rho; M)$, with a sufficient size $M$, can efficiently evaluate $\{o_i\}_{i=1}^H$. Specifically, $S(\rho; M)$ is divided into equally sized chunks, and multiple independent estimators are constructed. Subsequently, the approximations $\{\Tilde{o}_{i}\}_{i=1}^H$ of $\{o_i\}_{i=1}^H$ are obtained by using the median of means estimation~\cite{jerrum1986random}. The detailed procedure is summarized in Alg.~\ref{algorithm B1}.

The described procedure can be applied to various distributions of random unitary operators. One prominent example is tensor products of random single-qubit Clifford circuits. In this example, where each qubit is independently measured in a random Pauli basis, it can also be referred to as random Pauli measurements. The resulting classical shadow can be efficiently stored in classical memory using the stabilizer formalism. Furthermore, an interesting result demonstrates that $\mathcal{O}(\log(H)3^b/\epsilon^2)$ random Pauli measurements of $\rho$ suffice to evaluate $H$ bounded observables $\{O_i\}_{i=1}^H$ that are the tensor product of $b$ single-qubit observables. This evaluation guarantees $|\tilde{o}_i-o_i|\leq \epsilon$ with high probability for any $i\in[H]$. 

\end{document}